\documentclass[preprintnumbers, floatfix, letterpaper,aps,twocolumn,prd,epsfig,nofootinbib,natbib,longbibliography]{revtex4-1}
\usepackage{graphicx}

\usepackage{amsmath,amssymb,amsfonts,dcolumn,color,graphicx,graphics,latexsym}
\usepackage[mathscr]{eucal}
\usepackage[latin1]{inputenc}
\usepackage{enumerate}
\newcommand{\be}{\begin{equation}}
\newcommand{\ee}{\end{equation}}
\newcommand{\ba}{\begin{eqnarray}}
\newcommand{\ea}{\end{eqnarray}}

\newcommand{\lp}{l_{\mathrm{Pl}}}

\def\rcr{\rho_{_\mathrm{max}}^{^\mathrm{flat}}}

\def\lp{\ell_{\mathrm{Pl}}}
\def\be{\begin{equation}}
\def\ee{\end{equation}}

\begin{document}

\title{Hysteresis and beats in loop quantum cosmology}

\author{John L. Dupuy$^\star$ and Parampreet Singh$^\dagger$}
\affiliation{${}^\star$ Department of Physics and Astronomy, University of North Carolina at Chapel Hill, \\ Chapel Hill, North Carolina, USA \\ ${}^\dagger$ Department of Physics and Astronomy, Louisiana State University,\\
Baton Rouge, Louisiana, USA}

\begin{abstract}
Differences in pressure during expansion and contraction stages in cosmic evolution can result in a hysteresis-like phenomena in non-singular cyclic models sourced with scalar fields. 
We discuss this phenomena for spatially closed isotropic spacetime in loop quantum cosmology (LQC) for a quadratic and a cosh-like potential, with and without a negative cosmological constant using effective spacetime description of the underlying quantum geometry. 
Two inequivalent loop quantizations of this spacetime, one based on  holonomies of the Ashtekar-Barbero connection using closed loops, and another based on the connection operator, are discussed. Due to quantum geometric effects, both models avoid classical singularities, but unlike the holonomy based quantization, connection based quantization results in two quantum bounces. In spite of differences in non-singular effective dynamics of both the models, 
the phenomena of hysteresis is found to be robust for the $\phi^2$ potential.  
Quasi-periodic beats exist for a cosh-like potential, irrespective of the nature of classical recollapse whether by spatial curvature, or a negative cosmological constant. 
An interplay of negative cosmological constant and spatial curvature in presence of potentials results in rich features such as  
islands of cluster of bounces separated by accelerated expansion, and a universe which either undergoes a step like expansion with multiple turnarounds or quasi-periodic beats depending on a ``tuning'' of the steepness parameter of the potential.

\end{abstract}

\maketitle

\section{Introduction}
Scalar fields play an important role in cosmological dynamics. They are necessary for the inflationary paradigm and have been advocated as candidates for dark energy and dark matter. Apart from their thoroughly studied properties in phenomenology of our universe, they result in an interesting feature which becomes evident under suitable conditions in non-singular cyclic models. This is the phenomena of cosmological hysteresis \cite{st,sst}, first studied using a cyclic brane-world model with a time-like extra dimension \cite{sahni} and inclusion of a positive spatial curvature with and without a negative cosmological constant. With a suitably chosen scalar field potential,  resulting dynamics can result in a cyclic universe. A   lag in the scalar field trajectory in the expanding phase when compared to the contracting phase  results in cosmic hysteresis. Origin of this lag is easily evident for inflationary potentials.  In the expanding phase Hubble friction slows down the field resulting  in an accelerated expansion, but in the contracting phase, Hubble rate causes anti-friction resulting in the kinetic energy of the field to dominate and the inflaton behaves as a massless field. Since the pressure ($P$) of the scalar field during contraction is different than during expansion, there is an asymmetry in each cycle of recollapse-contraction-bounce-expansion-recollapse with $\oint P dV$ non-vanishing (where $V$ denotes the volume of the universe) \cite{nissim}. For sufficiently flat potentials such as $\phi^2$ inflation, pressure during expansion is less than during contraction and $\oint P dV$ is negative. The work done during contraction-expansion cycle leads to an increase in the size of the universe in the successive cycles occurring before inflation sets in. On the other hand, for steep potentials such as a  cosh-like potential, a candidate for dark matter models \cite{sw,matos}, $\oint P dV$ can be positive or negative, which can result in quasi-periodic beats of the scale factor of the universe \cite{st}. 

It is interesting to note that a cyclic universe undergoing hysteresis seems to possess an arrow of time even when dynamical equations are time reversible \cite{st,sst}. In cosmological dynamics one finds a point starting from which 
 the size of the universe increases in both temporal directions. This point can be viewed as the origin of the preferred direction of time. It has been argued that existence of this arrow of time is a result of attractor behavior of inflationary potentials where equation of state approaches to that of a cosmological constant during expansion and to of stiff matter during contraction \cite{sst}. This is in contrast to the conventional picture resulting from Tolman's model based on viscous fluids where the arrow of time is associated with entropy production \cite{tolman}.

The key to existence of phenomena such as hysteresis is the resolution of big bang/crunch singularities. This is difficult to achieve under generic conditions without understanding quantum gravity effects.  
It has been long expected that non-perturbative effects encoded in a quantum spacetime can resolve big bang/big crunch singularities resulting in a non-singular cyclic universe. In last 15 years, modifications to the physics of very early universe, resulting from loop quantum gravity (LQG), a candidate theory of 
non-perturbative background independent quantum gravity, have been extensively studied in loop quantum cosmology (LQC) \cite{as-status}, with the main result being the bounce of the universe in the Planck regime \cite{aps1,aps3}. Unlike various other approaches and models where singularity resolution generally requires exotic inputs or some fine tuning, resolution of strong cosmological singularities, such as the big bang and the big crunch is found to be generic  in the effective spacetime dynamics of LQC for various isotropic and anisotropic models \cite{ps09}. These results are found to be robust on inclusion of inhomogeneities in three-torus Gowdy cosmological model with linearly polarized
gravitational waves \cite{gowdy}. LQC models thus provide a robust platform to explore novel phenomenon arising from non-singular cyclic cosmological dynamics such as hysteresis or quasi-periodic beats.

Spatially closed models under suitable conditions undergo a recollapse of the universe resulting in a big crunch singularity in the future evolution. If the big bang/crunch singularities can be resolved, such as using quantum gravity effects, a non-singular cyclic model results. In LQC, quantization of $k=1$ model  has been rigorously performed and extensive numerical simulations confirm existence of a non-singular cyclic universe \cite{apsv,warsaw,cs-karami,cs-karami2}. Singularity resolution occurs in these models because of non-local quantum geometric effects which become significant in the Planck regime causing a bounce of the universe, thus avoiding big bang and big crunch singularities.  Two inequivalent quantizations of $k=1$ model exist in LQC. The first approach is based on using holonomies of the Ashtekar-Barbero connection to regularize field strength of the connection using closed loops \cite{apsv}. The second one is motivated from  constructions in loop quantization of anisotropic spacetimes and is based on the connection operator using holonomies over open loops \cite{cs-karami}.  A peculiar feature of the connection based quantization is the existence of two distinct quantum-turnarounds of the scale factor in dynamics \cite{cs-karami} which have been shown to exist for various types of matter content \cite{dupuy}. The two quantum bounces become indistinguishable in connection quantization if they occur at scale factors much larger than the Planck value. In this case both of the LQC models yield a cyclic cosmology with a single bounce followed by expansion and recollapse, and then a contraction and another bounce. It is important to note that in both of above models, LQG effects modify the Friedmann dynamics in a non-trivial way, especially the spatial curvature term. This is in contrast to the brane-world model where cosmological hysteresis was studied earlier \cite{sahni}, in which the modified Friedmann equation  has  resemblance to the one in $k=0$ model in LQC\footnote{Though there is no connection between LQC and brane-worlds, there is a curious similarity in the form of resulting modified Friedmann equations for spatially-flat models. Ignoring the bulk contributions to the brane, and replacing the brane-tension in the modified Friedmann equation on the spatially-flat brane in brane-world model with a time-like extra dimension \cite{sahni} with the bounce density in spatially-flat LQC \cite{aps3}, leads to the same modified Friedmann equation as in $k=0$ LQC. Given the complexity of quantum geometric effects on spatial curvature, this resemblance is no longer valid for spatially curved spacetimes.}, but, the spatial curvature has the same expression as in the classical theory.  Due to these reasons, whether or not hysteresis\footnote{For a discussion of conditions for existence of hysteresis in holonomy based quantization of spatially closed model in LQC, see Ref. \cite{cb}.}  and beats exist, and the way these phenomenon are modified in different models of LQC have been open questions.

Using effective dynamics, which has been rigorously shown to be valid in isotropic and anisotropic LQC \cite{numlsu-2,numlsu-4}, we carefully study in detail above two $k=1$ LQC models to understand the phenomena of hysteresis and beats. We find hysteresis to occur in both of the LQC models. 
Depending on the steepness parameter of the cosh-like potential, quasi-periodic beats are  found in both of the models of LQC.  Even slight differences in the initial conditions for the Ashtekar-Barbero connection in two LQC models quickly become significant in cyclic evolution resulting in differences in beating phenomena when the recollapse is determined by the positive spatial curvature.      
We consider separate cases where a negative cosmological constant is also present and plays a significant role in late time dynamics. This leads to various interesting features in dynamics which include step-like expansion with hysteresis in each step for $\phi^2$ potential, and a sensitive dependence of quasi-periodic beats on the steepness parameter of cosh-like potential. In presence of a negative cosmological constant, differences between two LQC models diminish. Some of the features we find in our analysis have been explored for the first time and were not reported in earlier studies. As an example, unlike earlier investigations for brane-world model \cite{st}, we find that quasi-periodic beats exist in LQC even when recollapse is sourced by the spatial curvature. 

The plan of the manuscript is as follows. In Sec. II we begin with a brief summary of the effective dynamics of holonomy and connection operator based LQC models for $k=1$ Friedmann-Lema\^itre-Robertson-Walker (FLRW) spacetime, which is followed by a discussion of  condition for hysteresis. In Sec. III LQC models are analyzed for $\phi^2$ potential with and without a negative cosmological constant. This is followed by an analysis of $\cosh(\phi)$ potential and quasi-periodic beats in Sec. IV. We summarize our results in Sec. V. \\

\section{Modified Friedmann dynamics}
Due to the underlying quantum geometry, the quantum Hamiltonian constraint in the loop quantization is a difference operator which can be approximated by an effective Hamiltonian derived using coherent state techniques \cite{vt}. Extensive numerical simulations show that quantum dynamics is captured extremely well by an effective dynamics derived from effective Hamiltonian for isotropic \cite{aps3,apsv,numlsu-2} as well as anisotropic models \cite{numlsu-4}. Our analysis will assume the validity of this effective dynamics. 
Quantum geometric effects enter into effective Hamiltonian in two ways. First from the field strength of the connection and second via the regularization of inverse volume terms. It turns out that the latter plays negligible role in singularity resolution of $k=1$ model\footnote{Note that inclusion of these further corrections do not affect singularity resolution, but only change the bound on energy density at the bounce \cite{cs-karami2,dupuy}.  Interestingly, a cyclic $k=1$ model in LQC can also be constructed solely using the inverse volume modifications too \cite{ps-at}.}  \cite{apsv}.
Hence, we will focus on effective dynamics resulting from field strength part of the constraint and neglect the inverse volume modifications.  In this section, we first summarize the effective Hamiltonian dynamics from two distinct quantizations of $k=1$ model: the quantization based on using holonomies over closed loops \cite{apsv}, referred to as holonomy quantization, and the connection based quantization \cite{cs-karami}. Both of these quantizations resolve the big bang and big crunch singularities but lead to phenomenologically different dynamics with the latter yielding two quantum turn-arounds instead of one for the former \cite{cs-karami,dupuy}. For detailed comparison of two approaches for various potentials, see Ref. \cite{dupuy}. Then we summarize the relationship between work done during contraction and expansion cycles and growth of the scale factor in hysteresis models.

\subsection{Effective dynamics from holonomy quantization}
In the loop quantization of homogeneous models, the Ashtekar-Barbero connection $A^i_a$ and its conjugate triad $E^a_i$ can be symmetry reduced to variables $c$ and $p$ respectively, which are the gravitational phase space variables. The physical 
volume of the unit sphere spatial manifold is $V = |p|^{3/2} = 2 \pi^2 a^3$, where $a$ denotes the scale factor of the universe. 
It is convenient to introduce the variable   $\beta := c|p|^{-1/2}$ which forms a canonical pair with $V$ and satisfies $\{\beta, V\} = 4 \pi G \gamma$. Here $\gamma$ is the Barbero-Immirzi parameter. As is conventional in LQC we will fix its  value as $\gamma \approx 0.2375$ using results from black hole thermodynamics in LQG \cite{gamma}. In the following, we will work with positive orientation of the triads thus eliminating the modulus sign. 

The effective Hamiltonian constraint for the holonomy quantization turns out to be \cite{apsv}:
\begin{widetext}
\begin{equation}\label{hcons1}
\mathcal{H}_{\rm{eff}}^{\rm{(hol)}} = \nonumber -\frac{3}{8 \pi G \gamma^2 \lambda^2} V [\sin^2(\lambda \beta - D) - \sin^2 D + (1+\gamma^2)D^2] 
~+ {\cal{H}}_{\mathrm{matt}} \approx 0~,
\end{equation}
\end{widetext}
where 
\be\label{Deq}
D := (\lambda (2\pi^2)^{1/3})/V^{1/3} ,
\ee
and $\lambda^2 = 4(\sqrt{3}\pi \gamma) \lp^2$ denotes the minimum area eigenvalue in LQG. Above ${\cal{H}}_{\mathrm{matt}}$ denotes the matter Hamiltonian.   Hamilton's equation for volume yields, 
\begin{equation}
\dot{V} = \{V,\mathcal{H}_{\rm{eff}}^{\rm{(hol)}}\} = \frac{3}{\gamma \lambda}V \sin(\lambda \beta - D)\cos(\lambda \beta - D).
\end{equation}
Using above equation with the vanishing of the effective Hamiltonian constraint $\mathcal{H}_{\rm{eff}}^{\rm{(hol)}} \approx 0$, we obtain the equation for the Hubble rate:
\begin{equation}\label{Hholonomy}
H^2 = \frac{\dot V^2}{9 V^2} = \frac{8\pi G}{3} (\rho - \rho_1 ) \left( 1-\frac{\rho - \rho_1}{\rcr}\right) .
\end{equation}
In the above equation,  
$\rcr = 3/(8 \pi G \gamma^2 \lambda^2)$ \cite{aps3}, and 
\begin{equation}
\rho_1 = \rcr[(1+\gamma^2)D^2 - \sin^2(D)]. 
\end{equation} 
Similarly, the Hamilton's equation for $\beta$ yields, 
\begin{equation}
\dot{\beta} = -4\pi G \gamma [ \rho - \rho_2 + P] 
\end{equation}
where $\rho_2$ is given by
\begin{equation}
\rho_2 = \frac{\rcr D}{3}[2(1+\gamma^2)D - \sin(2\lambda \beta - 2D)-\sin(2D)]
\end{equation}
and $P = -\frac{\partial\mathcal{H}_{\mathrm{matt}}}{\partial V}$ denotes the pressure of the matter component. 

In our analysis, the matter part of the Hamiltonian consists of a scalar field and a potential $U(\phi)$, in addition to a possible cosmological constant component. The Hamilton's equations for matter variables satisfy,
\begin{equation}
\dot{\phi} = \{\phi,  \mathcal{H}_{\rm{eff}}^{\rm{(hol)}}      \} = \frac{p_{\phi}}{p^{3/2}}, \; \; \; \dot{p}_{\phi} = - p^{3/2} \partial_{\phi} U(\phi)
\end{equation}
which result in the standard  Klein-Gordon equation for the field $\phi$ after taking time derivative $\ddot \phi$,
\begin{equation}
\ddot{\phi} + 3H \dot{\phi} + \partial_{\phi} U (\phi) = 0. 
\end{equation}

Using Hamilton's equations for gravitational and matter phase space variables, we can find physical solutions numerically. In our simulations, we impose initial conditions  on $\phi$, $\dot{\phi}$, and volume $V$, while the initial value of $\beta$ is determined using the vanishing of the effective Hamiltonian constraint. In particular, initial value $\beta_0$ satisfies 
\begin{eqnarray}\label{beta0a}
\sin^2(\lambda \beta_0 - D_0) &=& \nonumber  \rho_0 \left(\frac{8 \pi G \gamma^2 \lambda^2}{3} \right) + \sin^2(D_0) \\ && - (1 + \gamma^2) D_0^2 .
\end{eqnarray} 
Numerical solutions of these equations will be discussed in Sec. III and Sec. IV.

\subsection{Effective dynamics for connection operator quantization}
The connection operator quantization is an inequivalent quantization of $k=1$ FLRW spacetime obtained by using a different regularization of the quantum Hamiltonian constraint. Instead of using holonomies of the connection over closed loops to construct field strength of connection, one uses connection operator. Such a strategy is motivated from the loop quantization of anisotropic models and  
the resulting quantum dynamics is different from the one with holonomy quantization, resulting in two bounces instead of one which become prominent at small scale factors \cite{cs-karami}. For a detailed phenomenological comparison between effective dynamics of two LQC models, see \cite{dupuy}.

The connection operator quantization has a modified effective Hamiltonian constraint of the form \cite{cs-karami},
\begin{equation}
\mathcal{H}_{\text{eff}}^{\mathrm{(con)}} = \frac{-3}{8 \pi G \gamma^2 \lambda^2} V [ (\sin(\lambda \beta) - D)^2 + \gamma^2 D^2]+ \rho V ~.
\end{equation} 
Using Hamilton's equation, we get
\begin{equation}
\dot{V} = \frac{3}{\lambda \gamma}V \cos(\lambda \beta)[\sin(\lambda \beta) - D],
\end{equation}
which yields the following modified Friedmann equation \cite{dupuy}:
\begin{eqnarray} \label{Hconnection}
H^2 &=& \nonumber \frac{1}{\gamma^2	 \lambda^2} \cos^2(\lambda \beta)[\sin(\lambda \beta)-D]^2 \\
&=& \frac{8 \pi G}{3} (\rho - \rho_3 ) \left( 1-\frac{\rho - \rho_4}{\rcr}\right).
\end{eqnarray}
Here,
\be\label{rho34}
\rho_3 = \gamma^2 D^2 \rcr, 
\ee
and
\be
\rho_4 = D((1 + \gamma^2)D - 2\sin \lambda \beta) \rcr ~.
\ee
In contrast to the modified Friedmann equation for the holonomy quantization we find that there are two values of $\rho$ at which Hubble rate can vanish in the quantum regime which leads to two distinct quantum bounces. This occurs when $\sin(\lambda \beta) = \pm 1$: 
\be\label{conrho-bounce}
 {{\rho^\mp}} = \rcr((D \mp 1)^2 + \gamma^2 D^2) ~. 
\ee
Similarly, the time derivative of $\beta$ turns out to be 
\begin{equation}
\dot{\beta} = -4 \pi G \gamma(\rho - \rho_5 + P),
\end{equation}
with
\begin{equation}
\rho_5 = \frac{2 \rho_{\text{max}}D}{3}[(1+\gamma^2)D - \sin(\lambda \beta)].
\end{equation}
For the numerical simulations, we consider initial values $V_0,\phi_0, $ and $\dot{\phi}_0$, whereas $\beta_0$ is solved using the vanishing of the effective Hamiltonian constraint $\mathcal{H}_{\text{eff}}^{\mathrm{(con)}} \approx 0$ which yields 
\begin{equation}\label{beta0b}
\sin(\lambda \beta_0) = \left[\rho_0 \left(\frac{8 \pi G \gamma^2 \lambda^2}{3} \right) - \gamma^2 D_0^2\right]^{1/2} + D_0.
\end{equation} 
Due to the differences in Hamiltonian constraints for holonomy and connection quantization, the values for $\beta_0$ are slightly different even if all other initial conditions are set the same. In the numerical simulations discussed in this manuscript, this difference was of the order of 10$^{-3}$ to 10$^{-5}$.

\subsection{Condition for hysteresis}
We now briefly summarize the relationship between work done in a cycle of contraction and expansion in a non-singular closed universe and an increase in the scale factor of the universe in successive cycles. For discussion of hysteresis in cosmological setting, see \cite{nissim,st}, and Ref. \cite{cb} where LQC case based on holonomy quantization is also discussed. We consider the case when the recollapse is caused by the positive spatial curvature. 

At large scale factors,  $\rho/\rcr \ll 1$, and as a result the modified Friedmann equations in both LQC models approximate the classical Friedmann dynamics  for $k=1$ model:
\begin{equation}
 H^2 = \frac{8 \pi G}{3} \rho - \frac{1}{a^2}.
\end{equation}
The classical recollapse occurs when energy density and spatial curvature terms cancel, i.e. when $\rho = 3/(8 \pi G a^2)$. 
In the above approximation, the ``mass'' $M = \rho V$ contained in one cycle of the bouncing spatially-closed universe can be written in terms of the scale factor at the recollapse as $M = (3 \pi/4 G) a_{\mathrm{max}}$. Now consider a contraction-expansion cycle starting from recollapse point $a_{\mathrm{max}_{i}}$, undergoing a bounce, and expanding till consecutive recollapse point $a_{\mathrm{max}_{i+1}}$. If the value of maximum scale factor increases by $\delta a_{\mathrm{max}} := a_{\mathrm{max}_{i+1}} -a_{\mathrm{max}_{i}}$  between two such successive recollapse points, the mass $M$ increases by $\delta M$. This change in mass is related to the work done in each contraction-expansion cycle $\delta W = \oint P d V$ via the conservation law as $\delta W = - \delta M$. Thus one obtains \cite{st,nissim}, 
\begin{equation}
 \oint P d V = - \frac{3 \pi}{4 G} \delta a_{\mathrm{max}} ~.
\end{equation}
Whether or not there is hysteresis can thus be determined by 
plotting the equation of state $w = P/\rho$ versus the scale factor. 

In cyclic models such as of Ref. \cite{st}, it is also possible to easily understand hysteresis in terms of scale factors at the bounce by revisiting above argument for an expansion-contraction cycle. This is because the spatial curvature term in the cyclic brane-world model has the same form as the classical Friedmann equation.  The same is not possible for LQC models where quantum gravitational effects cause non-trivial modifications to this term which in particular become significant near the bounce. For models where there are modifications to spatial curvature terms, one may not even observe any increase in the bounce scale factors in successive cycles while at the same time finding an increase in maximum scale factors at recollapse points. In our analysis, we investigate hysteresis only via the recollapse points in the successive scale factors.  \\

\section{Chaotic inflation potential}
In this section we explore the dynamics in presence of chaotic inflationary models with scalar field potential  
\begin{equation}
U(\phi) = \frac{1}{2}m^2 \phi^2, 
\end{equation}
where $m$ is the mass parameter. We first explore hysteresis for cyclic models where the turnaround in the scale factor (or the volume) is generated by the spatial curvature. This is followed by the 
analysis in presence of a negative cosmological constant in Sec. IIIB.

\subsection{$k=1$ cyclic models with $\Lambda = 0$}
\begin{figure}[tbh!]
\centering
\includegraphics[scale=0.6]{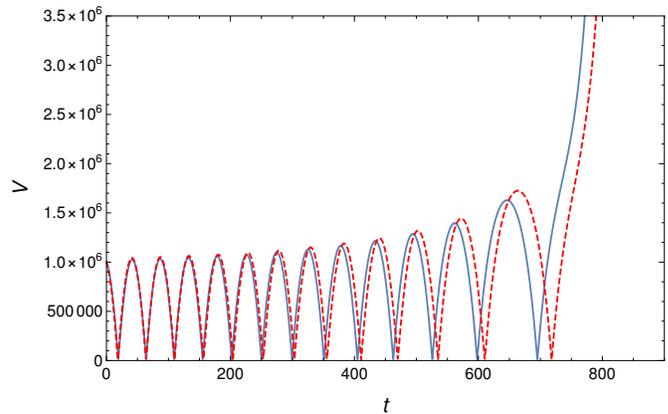}
\caption{Evolution of volume is plotted versus cosmic time in Planck units for two LQC models exhibiting cosmological hysteresis. The holonomy quantization is represented by the solid curve, and the connection quantization is represented by a dashed curve.   The initial conditions (in Planck units) for this simulation are $V(0) =10^2$, $\phi(0) = 3.0$, $\dot{\phi}(0) = 0.0135$, and $m=3\times 10^{-4}$.}
\label{fig1}
\end{figure}

Here we consider cases where cosmological turnaround in the scale factor of the universe is only due to the presence of positive spatial curvature. For this case, we consider a variety of initial conditions for both LQC models.  In Fig. \ref{fig1},  we show an example of a universe where the amplitude of the scale factor increases in consecutive cycles starting from $t = 0$ where the initial conditions are set. In this same example, the duration between cycles also increases in the forward evolution in positive time, which indicates cosmological hysteresis. As discussed earlier, for cosmological hysteresis to occur in cyclic models, the condition $\oint P \mathrm{d} V < 0$ is equivalent to a change in maximum value of the scale factor in the successive cycles to be positive.    

\begin{figure}[tbh!]
\includegraphics[scale=0.6]{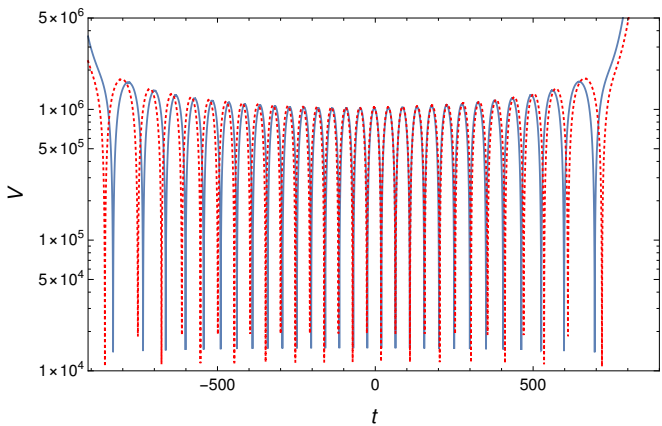}
\caption{Plot shows evolution discussed in Fig. \ref{fig1} for a longer time scale. Initial conditions are given at $t=0$ and hysteresis occurs in positive as well as negative time directions. Designations of particular models are the same as the previous figure. In connection based quantization, quantum bounces occur alternatively at different values of volume which correspond to two distinct quantum bounces in this model.}
\label{fig2}
\end{figure}

From Fig. \ref{fig1}, we can see that both for the holonomy quantized and connection quantized LQC models, the amplitude of scale factor of the universe increases in successive cycles starting from time $t = 0$. In addition, particularly at late times, the duration between  cycles  also increases. This is a clear demonstration of the cosmological hysteresis in LQC models caused by a lag in the expansion and contraction phases since the scalar field trajectory is not 
time reversed between expanding and contracting phases. Note that though both of the LQC models start from almost identical initial conditions (at $t = 0$) and their trajectories agree in the first few cycles, departures start emerging in subsequent cycles and become 
more pronounced in later cycles. In further time evolution, both of the LQC models result in an inflationary dynamics.  A comparison between these models for longer evolution is shown in Fig. \ref{fig2}, where see that the solutions are asymmetric about $t=0$. For the times less than $t=0$, the plot corresponds to a backward evolution in time starting from initial conditions at $t=0$. As earlier noted for the cyclic brane-world model \cite{st,sst}, hysteresis happens for both sides of $t=0$ in LQC models since the amplitude of cycles increases in successive cycles of evolution. Further, we can see existence of two distinct quantum bounces at alternating different values of volume in the connection based quantization of $k=1$ model in LQC.

It is to be noted that while above solutions display a clear evidence of hysteresis,  existence of this phenomena for the $\phi^2$ potential is subject to initial conditions because of the rapidity with which the solutions tend towards inflation. This observation is consistent with the attractor behavior in inflationary models in LQC \cite{inflation}. We explored a large range of initial conditions for different parameters and all these resulted in inflationary behavior after displaying few or some hysteresis cycles.  As can be seen from Fig. \ref{fig2}, for the considered initial conditions in the LQC models there are more than 10 cycles after $t=0$ before inflation starts.

\begin{figure}[tbh!]
\centering
\includegraphics[scale=0.6]{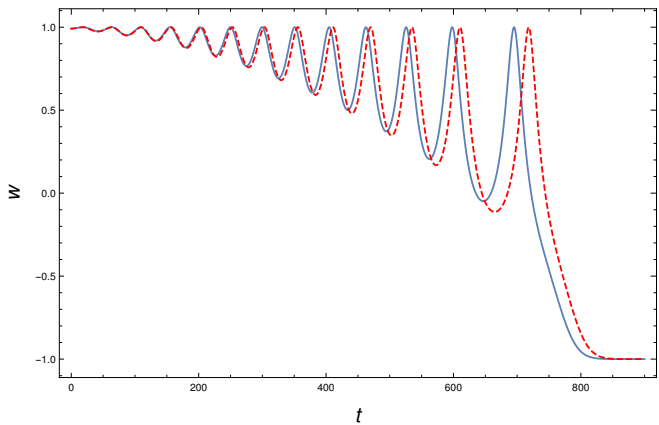}
\caption{Equation of state  is plotted versus time for the initial solutions shown in Figs. \ref{fig1} and \ref{fig2}. Designations of solutions are the same as those in the figures above.  }
\label{fig4}
\end{figure}

\begin{figure}[tbh!]
\centering
\includegraphics[scale=0.6]{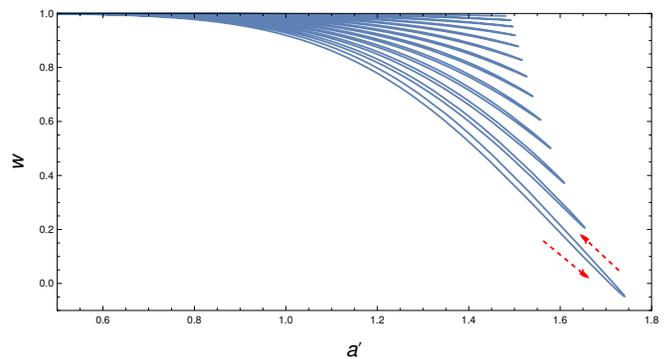}
\caption{The equation of state $w$ is plotted versus scale factor $a':= a/25$ for the holonomy solution shown in Figs. \ref{fig1} and \ref{fig2} for a part of the initial evolution starting from $t=0$.  The behavior in the connection quantization case is similar. 
Left (right) pointing arrow on one of the hysteresis loops in top plot shows collapsing (expanding) phase. }
\label{fig5}
\end{figure}

In these cycles of expansion and contraction, the equation of state initially varies around $w = 1$ and as the amplitude and duration between bounces grow, it approaches  $w = -1$ in the subsequent cycles. Eventually it reaches $w = -1$ when the inflationary phase begins. As can be seen from Fig. \ref{fig4}, the equation of state seems to mirror the behavior of the volume in Fig. \ref{fig2}. Note that the equation of state always becomes unity 
at each bounce. It reaches a local minimum as the volume reaches a local maximum in each cycle. This behavior is linked to hysteresis and can be understood by considering a plot of the equation of state versus the scale factor. From Fig. \ref{fig5}, we see that the path in the equation of state versus scale factor plot is different in the expansion and contraction phases. This process is repeated for each of the cycles of expansion and collapse and the result is a pattern of hysteresis-like loops. Though, it is much more subtle than the ideal case discussed in Ref. \cite{st}, each of these loops  enclose a small amount of area. This finite amount of area corresponds to a finite amount of work being done during each cycle. This work, in turn, drives the next cycle to a different value of maximum scale factor and a different duration. At later times, the equation of state is driven towards $w \approx -1$ where inflation sets in, causing hysteresis to die.

\subsection{$k=1$ cyclic models with $\Lambda < 0$} 

Presence of a negative cosmological constant can result in a recollapse of the scale factor in classical GR depending on the relative magnitude of the cosmological constant with respect to other matter densities. The recollapse is followed by a big crunch singularity in GR, which is avoided in LQC and a non-singular cycle results. Due to the behavior of energy density of cosmological constant, a large value of cosmological constant can dictate the recollapse even when spatial curvature is present.  In such a case, one expects that 
the differences between effective dynamics of holonomy and connection quantizations in LQC become negligible if the recollapses and bounces occur at volumes much greater than the Planck volume. This is because quantum geometric effects originating from intrinsic curvature play little role in dynamical evolution when volumes are large. And in such a case,  the volume at which two quantum bounces in connection based model occurs becomes virtually indistinguishable. 
On the other hand, if initial conditions are chosen where negative cosmological constant is not dominant then the situation is similar to the one discussed in the previous subsection. Therefore, in this part of our analysis we will consider a large value of negative cosmological constant.

\begin{figure}[tbh!]
\centering
\includegraphics[scale=0.6]{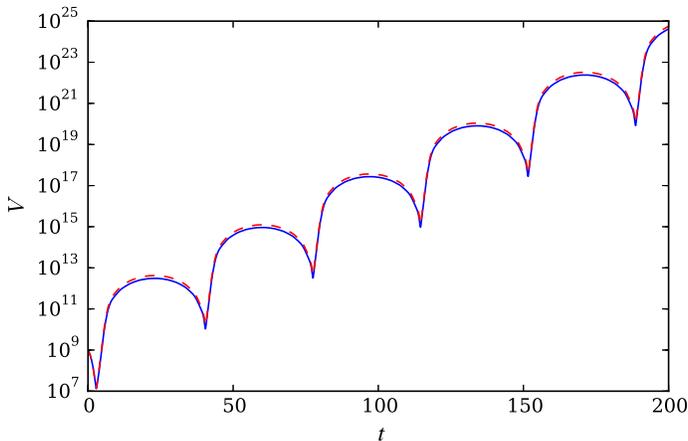}
\vskip0.3cm
\caption{Evolution is shown for  $\phi^2$ potential in a universe with a negative cosmological constant and positive spatial curvature. Plotted here are the holonomy solution (solid line) and the connection solution (dashed line).  The initial conditions selected for this solution are  $V(0)= 10^9$, $\phi(0)=10^{-7}$, $\dot{\phi}(0) = 0.1$, $m=1$, $\Lambda = -0.01$ (all in Planck units).}
\label{fig6}
\end{figure}

Let us first discuss the case of an ever increasing cyclic universe. Fig. \ref{fig6} displays results for such a universe, for the effective dynamics of holonomy and connection quantization in LQC, where initial conditions are set in the contracting phase.  We see that after a brief period of contraction, a quantum bounce occurs rather quickly which is followed by a short phase of accelerated expansion (evident 
by  a quick growth in the beginning of each cycle). In this phase the volume of the universe grows by over four orders of magnitude before the negative cosmological constant forces a turnaround. Though the scale factor increases between successive cycles, the duration of the cycles remains same. This plot confirms general expectations of hysteresis first discussed in Ref. \cite{st} for a cyclic model with a negative cosmological constant, and shows validity of those results for LQC. We also see that evolution in both the models is almost identical. For this reason, in presence of a negative cosmological constant we discuss only one of the models in the simulations. 

\begin{figure}[tbh!]
\centering
\includegraphics[scale=0.6]{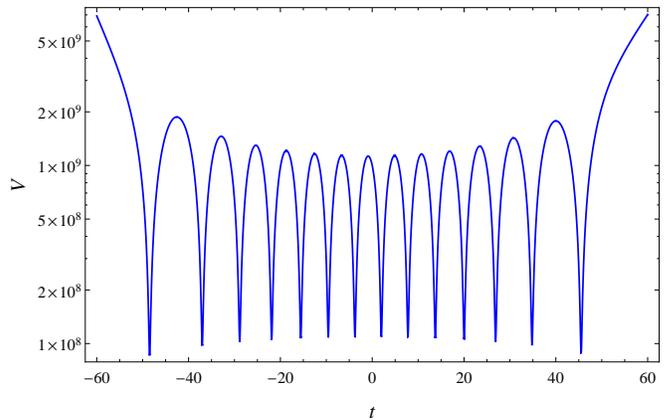}
\caption{Figure shows volume with respect to time for a universe subject to the initial conditions $V(0)= 10^9$, $\phi(0)=10^{-7}$, $\dot{\phi}(0) = 0.1$, $m=0.01$, $\Lambda = -0.01$. Note the difference in comparison to Fig. \ref{fig6} due to the change in the value of $m$. Only the holonomy solution is plotted because the dynamics from connection quantization agrees approximately with the solution presented here.}
\label{fig7}
\end{figure}

\begin{figure}[tbh!]
\centering
 \includegraphics[scale=0.6]{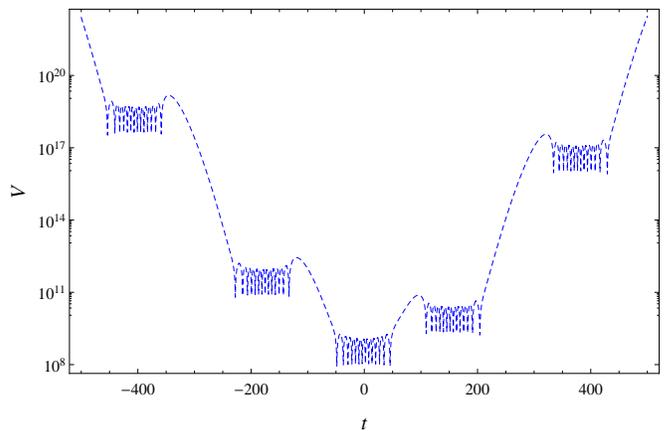}
 \caption{Longer time evolution of a universe with the same initial conditions as the solution presented in Fig. \ref{fig7}. Only the effective dynamics solution from holonomy quantization is shown. Islands of cluster of bounces separated by short but rapid expansion phases are visible.}
\label{fig8}
 \end{figure}

The simulation presented in Fig. \ref{fig7} has some resemblance to the one in Fig. \ref{fig2}. We see that the scale factor and the duration between cycles increases in successive cycles, before inflation takes over dynamics.   
However, because of the large negative cosmological constant, inflation does not last long. Instead something dramatic occurs which is evident in Fig. \ref{fig8} depicting late time evolution of Fig. \ref{fig7}. 
The negative cosmological constant causes a turn-around and as a result the universe inflates only for some time before getting stuck in a cluster of bounces. Each of these have a form similar to those shown in Fig. \ref{fig7}. The negative cosmological 
constant thus not only halts inflation but also forces the scale factor into these clusters of bounces or ``bouncing islands"  until a short period of rapid expansion  occurs again. As with the case in Fig. \ref{fig2}, we see that the solution is time asymmetric.  The cycle of short inflation and occurrence of cluster of bounces continues in subsequent evolution. \\

\section{Cosh-like Potential}
We now discuss cyclic models for $k=1$ LQC for the case of a $\cosh$-like potential and compare the solutions in effective dynamics of holonomy and connection quantizations. The $\cosh$-like potential is sometimes used for scalar field candidates of dark matter \cite{sw,matos}, and is of the form 
\begin{equation}\label{cosh-pot}
U(\phi) = m^2 [\cosh(q\phi/\tilde{m}_P) - 1] .
\end{equation}
Here $q$ is the steepness parameter of the potential and $\tilde{m}_P = 1/\sqrt{8 \pi G}$. We will set  $m=10^{-6}$ and steepness parameter will be varied from $q=1.0$ to $q=7.0$. As discussed in Ref. \cite{st},  the value 
of the parameter $q$ greatly affects the behavior of the solutions and for such values of steepness parameter inflation may not occur. Unlike the case of $\phi^2$ potential, the sign of $\oint P \mathrm{d} V$ can be positive as well as negative, resulting in quasi-periodic beats \cite{st}.   In the following, we first explore the presence and the lack of cosmological quasi-periodic beats in the cases where a recollapse is caused by positive spatial curvature only. We  then consider the cyclic model with a positive spatial curvature and a negative cosmological constant.   In contrast to earlier results \cite{st}, we find beats to exist for these models even when recollapse occurs due to spatial curvature.

\begin{figure}[tbh!]
\centering
\includegraphics[scale=0.6]{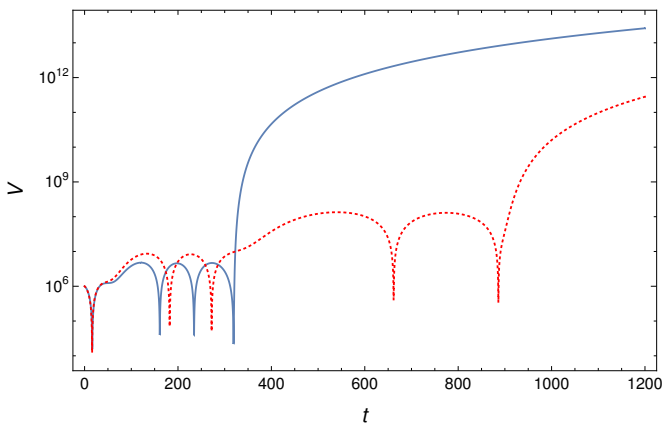}
\caption{Evolution of volume of the  universe with  a $\cosh(\phi)$-like potential with steepness parameter $q=1.5$. The holonomy solution is represented by a solid line and the connection solution by a dotted line. The initial conditions for these solutions were $V(0)=10^6$, $\dot{\phi}(0)=0.0141$, $\phi(0) = 0.5$ (in Planck units).    }
\label{fig10}
\end{figure}

\subsection{$k=1$ cyclic models with $\Lambda = 0$}   

In presence of positive spatial curvature for the above potential there can be a recollapse of the scale factor in the classical regime causing a big crunch singularity in the future. In LQC, quantum gravitational effects avoid both the past big bang and the future big crunch singularities causing a cyclic evolution. The first case we discuss below is an example where the holonomy quantization and the  connection operator quantization give radically different solutions for similar initial conditions. Note that the initial conditions are quite similar but not the same for the holonomy and connection quantizations because their effective Hamiltonians differ and lead to slightly different values of $\beta_0$ (see eqs.(\ref{beta0a}) and (\ref{beta0b})). %
From Fig. \ref{fig10}, one can see that the selected initial conditions result in a universe that is very quickly contracting. A bounce occurs and the positive spatial curvature in the dynamics of holonomy and connection quantizations almost leads to another successive turnaround  at $t\approx 75$ if not for the steep potential. The potential causes the volume to grow until spatial curvature results in a  recollapse. The universe bounces at irregular values of volume until potential starts dominating the dynamics and an accelerated expansion occurs. 
It is to be noted that for the simulation shown in Fig. \ref{fig10}, different models result in solutions  which are radically different after only a short amount of time.  For these simulations, given the relatively small value of  steepness parameter, there are no quasi-periodic beats. These arise for larger values of $q$ which is discussed below.

\begin{figure}[tbh!]
\centering
\includegraphics[scale=0.6]{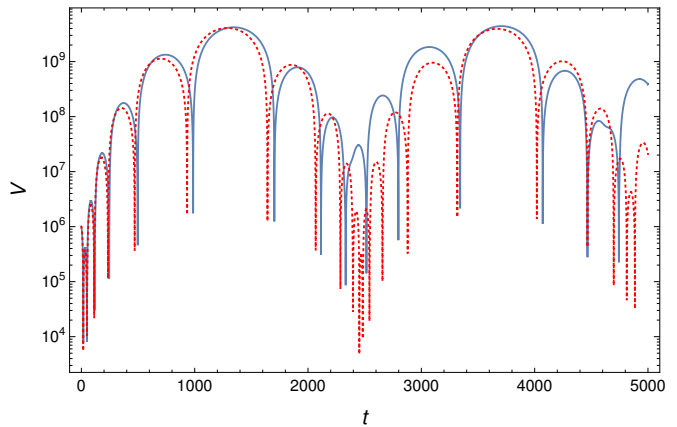}
\caption{Evolution in  a quasi-periodically beating universe. The holonomy solution (solid) and connection solution (dotted) are shown for $q=4.0$ with $V_0 = 10^6$, $\dot{\phi}_0=0.0141$, and $\phi_0 = 0.5$ (in Planck units).}
\label{fig11}
\end{figure}

\begin{figure}[tbh!]
\centering
\includegraphics[scale=0.6]{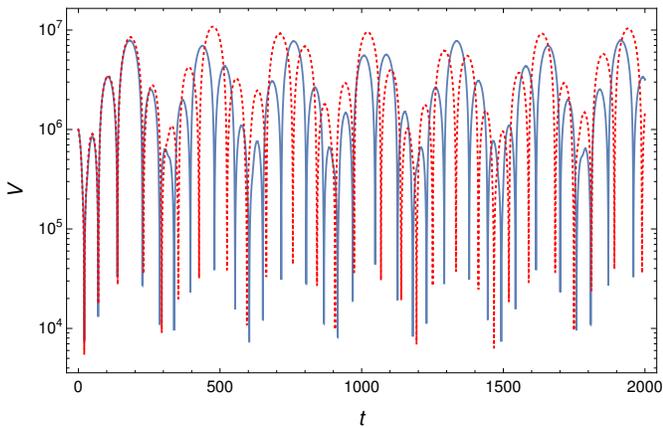}
\caption{Evolution  for a universe subject to the same initial conditions as Fig. \ref{fig11}, except that $q=5.5$. The solution from  holonomy quantization is shown by solid curve, while from connection quantization is shown by dotted curve.} 
\label{fig12}
\end{figure}

Fig. \ref{fig11} shows the emergence of quasi-periodic beats which occur because of  variations in the hysteresis loop. In the previous study based on cyclic brane-world model, quasi-periodic beats were found to be absent in this case, and instead a  stochastic behavior was found \cite{st}. 
Phenomena of beats appear and disappear as steepness parameter is varied. Beats appear as the steepness parameter, $q$, is increased. They emerge around $q=3.5$ and subsequently become less regular as $q$ is increased to values larger than $q = 7.0$. We find that though quasi-periodic beats appear in both of the LQC models, there are some major deviations between the models. An example is shown for the case of $q = 4.0$ in Fig. \ref{fig11}, where we see that the structure of expansion and contraction cycles does not exactly repeats itself and is quasi-periodic. Another example is shown for the case of $q = 5.5$ in Fig. \ref{fig12}. The quasi-periodic structure is evident in both the holonomy as well as connection quantization solutions, with a period of around $T\approx 300$.

\begin{figure}[tbh!]
\includegraphics[scale=0.6]{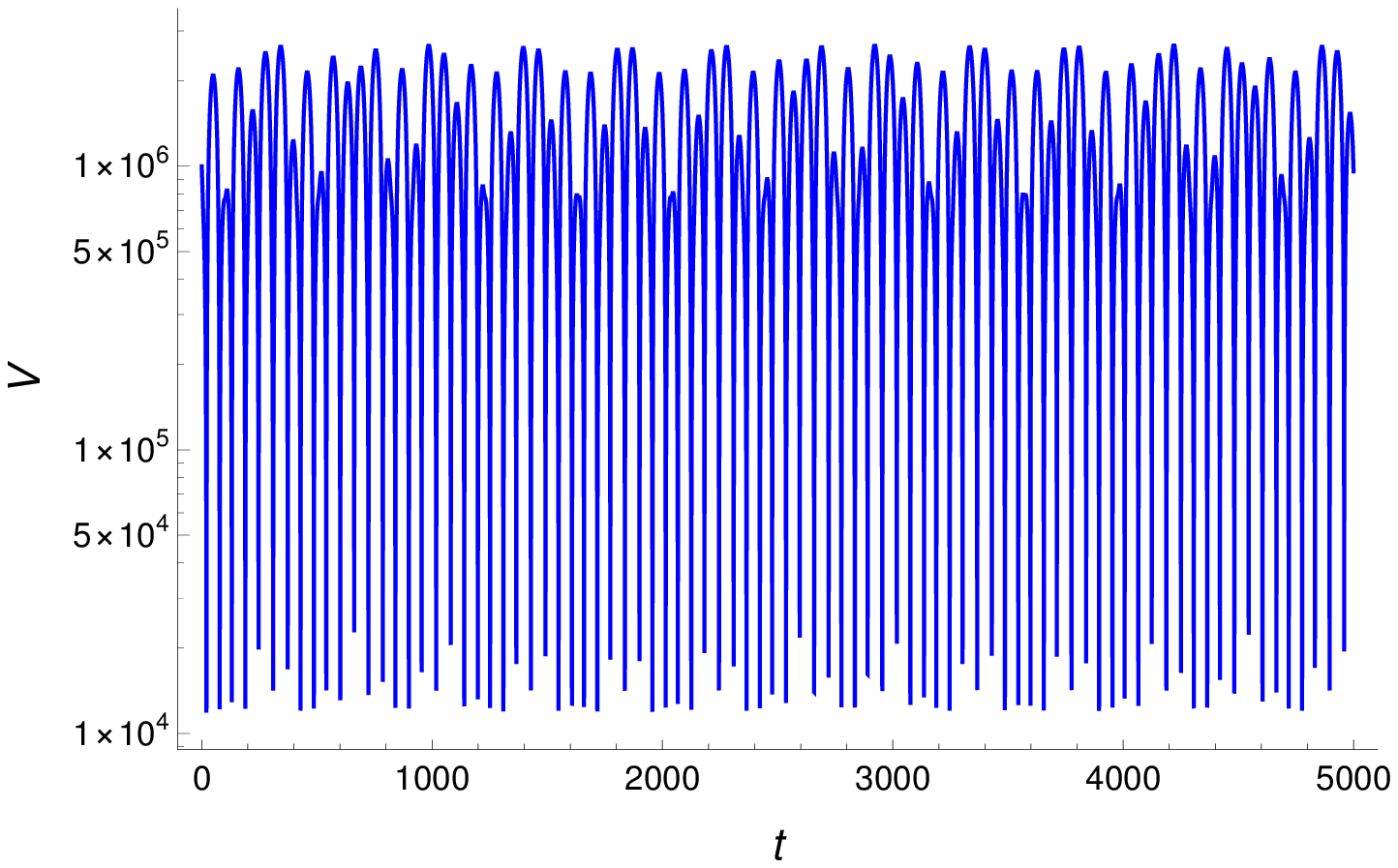}
\includegraphics[scale=0.6]{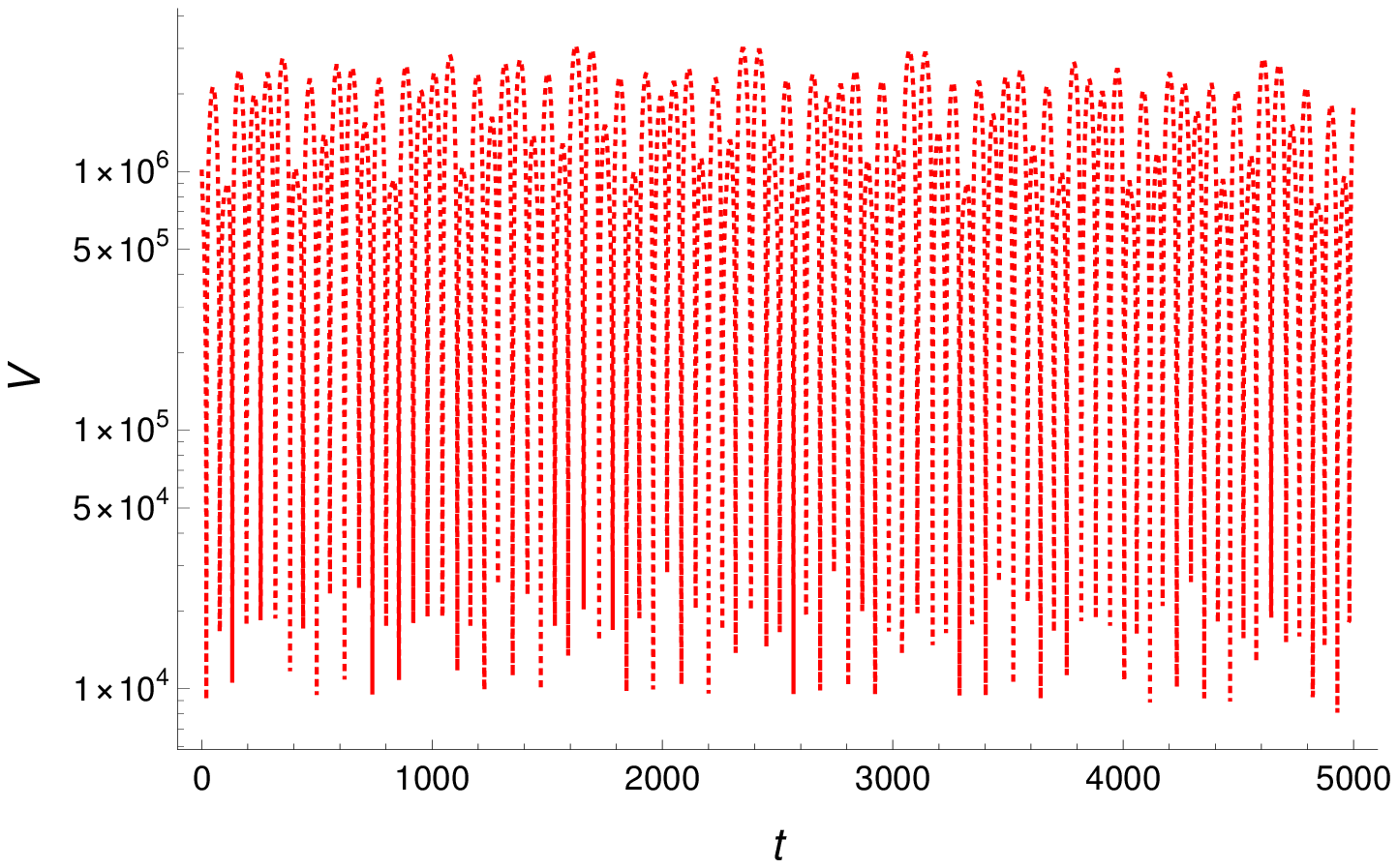}
\caption{Behavior of the volume is plotted for holonomy (top plot) and connection quantization (bottom plot) for $q=7.0$. Initial conditions are the same as Fig. \ref{fig11} except for the change in the value of steepness parameter.}
\label{fig13}
\end{figure}

Fig. \ref{fig13} shows the case of $q=7.0$ where the quasi-periodic structure in the effective dynamics of holonomy and connection quantizations is disappearing.  For $q > 7.0$, the periodic structure seems to give way to more stochastic behavior in both of  the models.

\begin{figure}[tbh!]
\centering
\includegraphics[scale=0.6]{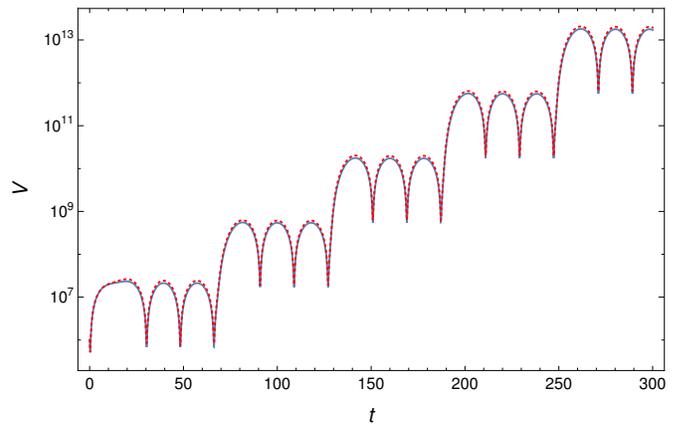}
\caption{Evolution  for a universe subject to a $\cosh$-like potential with steepness parameter $q=2.0$ with a negative cosmological constant. The holonomy quantization solution (solid curve) and connection quantization solution (dotted curve) agree with each other. Initial conditions for this solution are: $V_0 = 10^6$, $\phi_0=1.03$, $\dot{\phi}_0=0.5$, with $\Lambda=-0.01$ (in Planck units). }
\label{fig14}
\end{figure}

\subsection{$k=1$ cyclic models with $\Lambda < 0$}

In the case of a recollapse sourced by a negative cosmological constant in presence of positive spatial curvature, the pattern of beats become more regular than in the case when only spatial curvature is present. Earlier investigation for this case for the brane-world cyclic model showed 
that as the steepness parameter is increased the behavior of the cyclic universe changes dramatically \cite{st}. Our investigation for LQC models confirms that moderately small values of the steepness parameter  gives rise to a steady increase in the amplitude of the scale factor. Our analysis also reveals some new features. Let us start with discussion of simulation shown in Fig. \ref{fig14}, where we show the solutions for holonomy and connection quantizations for a universe  with a steepness parameter of $q=2.0$ for potential \eqref{cosh-pot}. Comparing to the case of $\phi^2$ potential in Fig. \ref{fig6}, we see some similarities but also some differences. As in the case of $\phi^2$ potential with a negative cosmological constant, the holonomy and connection quantizations lead to solutions which are almost identical. Although, there are no 
quasi-periodic beats in this particular case of the chosen value of $q$, the universe undergoes an interesting expansion dynamics. The cosmic evolution has a step-like behavior with each step resulting in multiple cycles with almost same expansion factors.  After these multiple cycles, which turn out to be three in the presented case, inflation attempts to take over turnaround and the scale factor increases to a higher amplitude. However, the negative cosmological constant does not allow inflation forcing another turnaround and the cycle repeats at a higher volume. In comparison to the phenomena of multiple bounces seen for $\phi^2$ potential in Fig. \ref{fig7} and \ref{fig8}, the effect is subdued in the present case.

\begin{figure}[tbh!]
\centering
\includegraphics[scale=0.4]{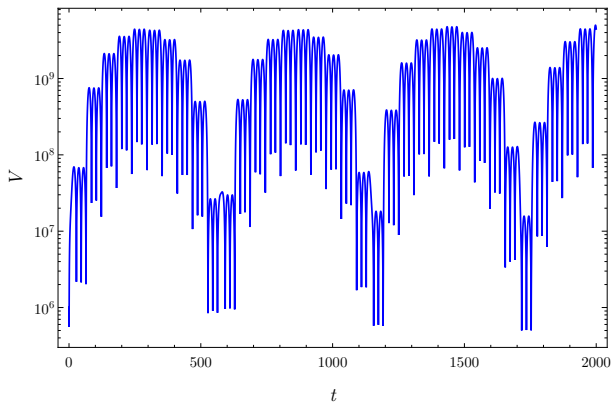}
\caption{Behavior of the volume versus time  for a universe with the same initial conditions as Fig. \ref{fig14}, with $q=2.205$. Emergence of quasi-periodic beats is evident.}
\label{fig15}
\end{figure}

\begin{figure}[tbh!]
\centering
\includegraphics[scale=0.4]{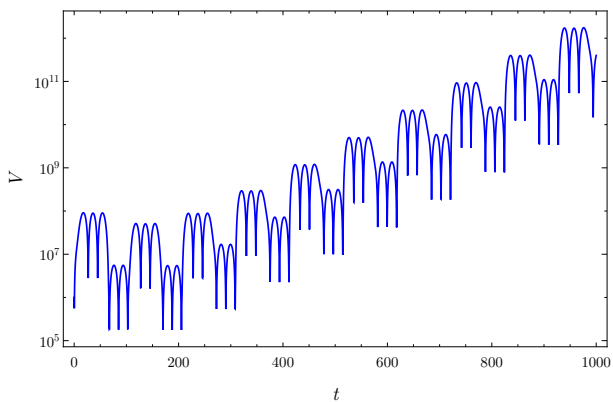}
\caption{For the steepness parameter $q=2.3$, the behavior of volume in time is shown. Initial conditions are same as Fig. \ref{fig14}. With a small change in steepness parameter in 
contrast to Fig. \ref{fig15}, the beats disappear. }
\label{fig16}
\end{figure}

\begin{figure}[tbh!]
\centering
\includegraphics[scale=0.4]{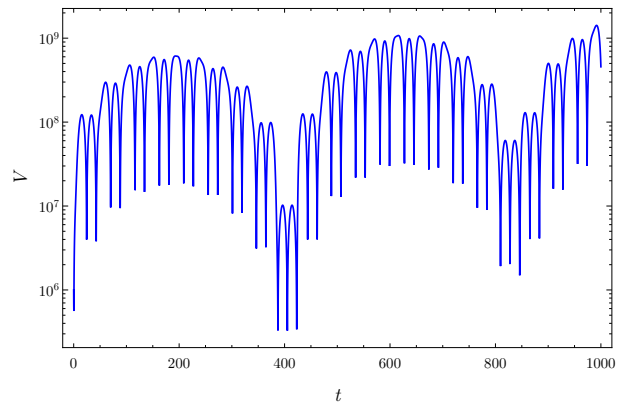}
\caption{Evolution of volume versus time  is shown for the case $q=2.5$ with the initial conditions  same as in Fig. \ref{fig14}. The re-emergence of beats is to be noted.}
\label{fig17}
\end{figure}

Interestingly, as we vary steepness parameter $q$ from 2.0 to 2.5 in small increments we see the appearance and disappearance of the quasi-periodic beats. 
We found  that such a  behavior is more regular for recollapse caused by negative cosmological constant than for the cases when the recollapse is caused by the positive spatial curvature. 
This behavior is shown in the next three figures which correspond to $q=$ 2.205 (Fig. \ref{fig15}), 2.3 (Fig. \ref{fig16}), and 2.5 (Fig. \ref{fig17}). Solutions in holonomy and connection quantization are almost the same, hence we only show the solutions from effective dynamics corresponding to holonomy quantization.  We see that though quasi-periodic beats are not present for  $q=2.0$, but they emerge when we increase $q$ to 2.205. As we increase $q$ to 2.3, the beat structure disappears. The cyclic evolution becomes more interesting than in the case of Fig. \ref{fig14}. Now the steps in the evolution are richer and the expansion is not monotonic. A further increase of $q$ to 2.5 causes the quasi-periodic beat structure to emerge once again as is shown in Fig. \ref{fig17}. If $q$ is increased to 3.0, the beats structure disappears.

\begin{figure}[tbh!]
\centering
\includegraphics[scale=0.4]{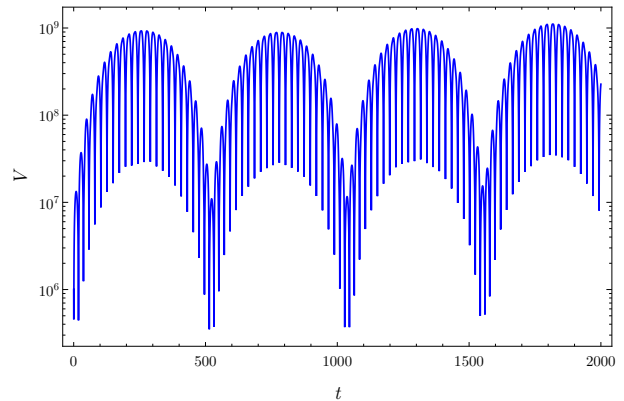}
\caption{Variation of volume with time on a log-linear scale for a universe with the same initial conditions as Fig. \ref{fig14}, except that a steepness parameter of $q=5.0$ was assigned. The evolution consists of large cycles, each of the cycles composed of quantum beats. }
\label{fig18}
\end{figure}

A case of higher value of steepness parameter is shown in Fig. \ref{fig18} where $q=5.0$. As before we show only  the holonomy case, as the connection based quantization yields similar  dynamics. The figure shows the beats phenomena nested  within larger cycles. The regularity of the larger cycles is quite evident, especially when compared to the beats patterns in universes where turnaround was caused only by positive spatial curvature. Analyzing this solution for longer time scales confirms that both the period and the amplitude of these large cycles stay approximately constant.

\begin{figure}[tbh!]
\centering
\includegraphics[scale=0.4]{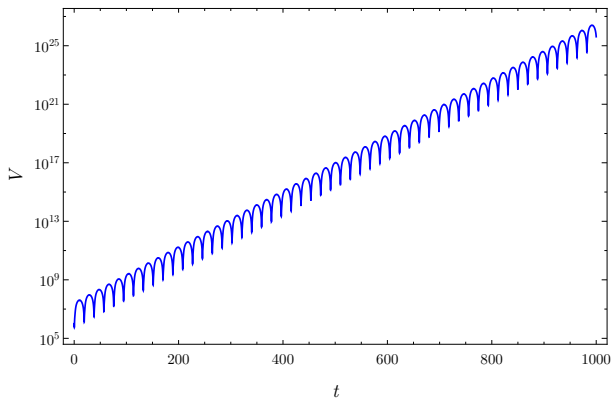}
\caption{Evolution is shown the same initial conditions as Fig. \ref{fig14} with $q=6.0$. Quasi-periodic beats are replaced by multiple ever increasing cycles. }
\label{fig19}
\end{figure}

Another case of a higher value of steepness parameter is shown in Fig. \ref{fig19} for $q=6.0$ for the holonomy case. Unlike the case of $q=5.0$ we find that the quasi-periodic beats have completely disappeared. Rather the universe undergoes an expansion phase with multiple bounces at ever increasing values of scale factor. This figure has some similarity with the case of $\phi^2$ potential discussed in Fig. \ref{fig6}. For higher values of steepness parameter, we find the beats phenomena to become less regular to occur.
We find that the existence of quasi-periodic beats is sensitive to the value of steepness parameters if other initial conditions are not changed. In certain cases evolution has close similarities to the $\phi^2$ potential. And, in some of the 
cases very interesting step like expansion behavior with multiple cycles in each step emerges. As with the case of the $\phi^2$ potential, the presence of negative cosmological constant mitigates differences between two LQC models.

\section{Conclusions}
In the presence of scalar fields the dynamics of cyclic models can be quite interesting. One of such phenomena is the presence of hysteresis arising due to differences in the equation of state in the expanding and the contracting branches \cite{nissim,st,sst}. Even with dynamical equations respecting symmetry of time reversal, hysteresis seems to bring out an arrow of time in such an evolution. The goal of this manuscript was to explore the phenomena of hysteresis in two different quantizations of $k=1$ model in LQC. The first loop quantization  arises from considering holonomies of the Ashtekar-Barbero connection over closed loops \cite{apsv,warsaw}, and the second one results from considering connection operator \cite{cs-karami}.
    We studied above three models for $\phi^2$ inflationary potential, and $\cosh$-like potential a candidate of cold dark matter. Dynamics in both potentials was explored with classical turnaround forced by positive curvature as well as by a negative cosmological constant.

For the case of $\phi^2$ potential, we find that both of the LQC models exhibit hysteresis robustly.  We find a clear evidence of hysteresis in terms of the increase in the scale factor at recollapse in the subsequent cycles.  In presence of a negative  cosmological constant, differences between cosmological dynamics from both of the LQC models diminish. For certain initial conditions, we find a novel phenomena of period of hysteresis followed by step-like rapid growth of scale factor. This is found for both the models and is a result of a competition between inflation and recollapse caused by a negative cosmological constant. The latter does not let the universe enter an inflationary regime but rather traps it for a period into a cluster of bounces and recollapses. In the ensuing hysteresis, conditions become favorable for a rapid expansion which is soon followed by another cluster of bounces but now occurring at a higher value of scale factor. This interesting behavior repeats itself in the subsequent evolution.  

In the case of $\cosh$-like potential (\ref{cosh-pot}), depending upon the value of steepness parameter, there can be an increase or decrease in the maximum value of scale factor in subsequent cycles. This results in possibility of quasi-periodic beats which were found earlier for the cyclic brane-world model but only when recollapse is sourced by a negative cosmological constant \cite{sahni}. In our analysis, we find existence of beats structure  even in absence of negative cosmological constant. Unlike the case of $\phi^2$ potential, when recollapse is caused by spatial curvature, differences between both of the models are most pronounced for this potential. Tiny differences in the initial conditions for connection $\beta$ caused by differences in the Hamiltonian constraint of two LQC models leads to significant departures in dynamical evolution within a short period of time when recollapse is caused by spatial curvature. Even though the qualitative properties of the solutions are similar but the ``beats" patterns for each model get gradually more out of phase. 
 For certain higher values of the steepness parameter, quasi-periodic beats become less regular. 

In presence of negative cosmological constant, $\cosh$-like potential results in a more regular existence of beats in both of the LQC models. Unlike the case when recollapse is caused by spatial curvature, in the present case a very interesting phenomena of presence and absence of beats appears depending on the choice of the steepness parameter. In particular, when the steepness parameter is varied in small steps in a range, we find the universe undergoing and going out of the beats phenomena depending on the value of the steepness parameter. In cases where quasi-periodic beats are absent, expansion of the universe undergoes step-like expansion, or step-up step-down like expansion with multiple cycles in each step. Whether the universe undergoes quasi-periodic beats or novel expansion as above is sensitive to small changes in the steepness parameter. In presence of negative cosmological constant, the steepness parameter thus serves as a ``tuning'' to select  a particular ``beating'' or its absence for the universe.

In summary, hysteresis and beats are found to be robust for effective dynamics of both holonomy and connection quantization of spatially closed FLRW model in LQC. By and large, the two studied potentials lead to vastly different universes with interesting phenomena. Beating universe scenarios are absent from the $\phi^2$ potential, whereas they are a very typical phenomenon in the $\cosh$-like potential depending on the steepness parameter.  Differences between different models are more apparent in the case of $\cosh$-like potential and when classical recollapse is caused by spatial curvature. There are various avenues to explore genericness of these phenomenon in LQC. One of them is anisotropic spacetimes where there exist inequivalent quantizations with non-singular cosmic evolution. As in the case of the spatial curvature, quantum geometric effects modify the the anisotropic shear in a non-trivial way. It will be interesting to understand the way anisotropies effect the occurrence of hysteresis and beats in LQC models.

\section*{Acknowledgements} 
We thank Varun Sahni and Alexey Toporensky for discussions, and Bao-Fei Li for discussions and comments.  J.D. thanks the REU program in Physics and Astronomy at LSU during
which a large part of this work was completed. P.S. is supported by NSF grant PHY-1454832.

\end{document}